# The Fast and Accurate Approach to Detection and Segmentation of Melanoma Skin Cancer using Fine-tuned Yolov3 and SegNet Based on Deep Transfer Learning

M. Taghizadeh, K. Mohammadi

*Abstract*— Melanoma is one of the most serious skin cancers that can occur in any part of the human skin. Early diagnosis of melanoma lesions will significantly increase their chances of being cured. Improving melanoma segmentation will help doctors or surgical robots remove the lesion more accurately from body parts. Recently, the learning-based segmentation methods achieved desired results in image segmentation compared to traditional algorithms. This study proposes a new approach to improve melanoma skin lesions detection and segmentation by defining a two-step pipeline based on deep learning models. Our methods were evaluated on ISIC 2018 (Skin Lesion Analysis Towards Melanoma Detection Challenge Dataset) well-known dataset. The proposed methods consist of two main parts for real-time detection of lesion location and segmentation. In the detection section, the location of the skin lesion is precisely detected by the fine-tuned You Only Look Once version 3 (F-YOLOv3) and then fed into the fine-tuned Segmentation Network (F-SegNet). Skin lesion localization helps to reduce the unnecessary calculation of whole images for segmentation. The results show that our proposed F-YOLOv3 performs better at 96% in mean Average Precision (mAP). Compared to state-of-the-art segmentation approaches, our F-SegNet achieves higher performance with 95.16% accuracy.

## I. Introduction

Skin cancer is one of the most widely diagnosed cancers globally, and it is more likely to develop and spread. When melanocyte cells proliferate and are out of control, they cause a malignant melanoma skin tumor. The fatality of malignant melanoma from various skin tumors is indicated by a higher mortality rate and susceptibility to spreading into surrounding tissues [1]. The number of diagnosed skin cancers caused by this cancer is estimated at 104,450 and 11,650 deaths in the United States in 2019, except for squamous cell and basal cell skin cancers [2]. Compared to all types of skin tumors, melanoma represents 92.5% of cases and 62.1% of deaths from this cancer. Early diagnosis of skin lesion cancer, on the other hand, is treatable and can lower melanoma mortality rates. Health check-ups of skin cancers by comparing skin lesions and natural tissue through the naked eye can lead to misdiagnosis. The most trustworthy imaging methodology for skin lesions is dermoscopy. Filtering the reflection from the surface of the skin or reducing the reflection is the main point in dermoscopy to create a magnified, high-resolution image. Given that dermatologists have challenges in improving the diagnosis of skin cancer further, better visualization of skin lesions through dermoscopic imaging and thus improved sensitivity leads to a more accurate diagnosis of skin cancer, and features that lead to the proper diagnosis of non-cancerous suspicious lesions can be improved in comparison to visual inspection. Due to the complexity, time-consuming, fault-prone, and the diagnosed results may be subjective and have different results in manual inspection [3]. A trustworthy and automatic computer-aided diagnosis (CAD) system is a valuable assessment tool to support dermatologists' decisions for skin cancer. The segmentation of melanoma lesions is critical in creating an automated melanoma identification system to improve diagnostic performance.

There are two types of automated melanoma segmentation methods called traditional and deep learning techniques. Adaptive thresholding [4], level set segmentation [5], iterative selection threshold [6, 7], iterative stochastic area merging [8], and Otsu's thresholding [9] are examples of conventional approaches for melanoma segmentation. Clinical or natural artifacts, on the other hand, impair the performance of threshold-based segmentation methods [4, 6, 8, 9]. Otsu's thresholding [9] provides acceptable segmentation performance; nevertheless, the borders of the segmented region are uneven, reducing picture quality. The restriction of Otsu's thresholding was solved in [10] by averaging the intensities throughout the pixel. Some methods use an object identification algorithm on dermoscopic pictures to identify melanoma-affected areas in superior segmentation results [11, 12]. The hyper-graph was utilized in [11] to map the saliency of melanoma regions using superpixel data. Deep learning approaches [13-16] for melanoma segmentation have demonstrated considerable performance gains over standard segmentation methods.

This study suggests an incorporated two-phase diagnostic approach and segmentation based on a convolutional neural network (CNN) to overcome the computational overhead and precisely detect and segment skin lesion localization. In order to improve the diagnosis efficiency, it is an essential step for detection and melanoma lesions segmentation. Firstly, we detect the skin lesion regions using our fine-tuned You Only

Look Once version 3 network (F-YOLOv3) for melanoma lesion localization as a target area. Then we utilize the extracted regions to feed into our fine-tuned segmentation network (F-SegNet). This work focuses on detection and segmentation over the ISIC 2018 dataset, in which the training data consists of 2594 images and corresponding ground truth masks. The results and method of this research can be beneficial for using deep learning architectures to diagnose skin lesions.

The primary purpose of this paper is to supply a trustworthy system according to deep learning with extremely accurate detection and segmentation of skin lesions for the diagnosis of skin disease.

We suggest a two-phase procedure for melanoma segmentation to dominate the computational overhead: the composition of melanoma localization and segmentation steps. At first, we applied F-YOLOv3 to determine the location of the melanoma skin lesion as an area of interest. After determining the location of melanoma, F-SegNet is used for melanoma-affected lesion segmentation. What we have done in this paper are:

1) Our F-YOLOv3 method can localize several melanoma skin lesions that exist in a single image.
2) Based on a deep convolutional-based network, F-YOLOv3 is trained for the localization of the lesion region in the image.
3) A real-time lesion localization approach has been suggested to feed extracted regions into F-SegNet.
4) Due to the use of F-YOLOv3, it is not necessary to use an additional step to remove artifacts such as black frames, color charts, and clinical rule marks.
5) The melanoma lesion is accurately segmented by our F-SegNet and is evaluated on the ISIC 2018 dataset.

II. RELATED WORK

Several studies have been performed to prepare CAD systems for diagnosing melanoma skin lesions. Problems such as many differences between classes of many lesions types in terms of low contrast, texture [17], size, color [18], shape [19], and the presence of artifacts in dermoscopy images cause many difficulties and, therefore, correctly recognize of melanoma skin lesions from non-melanoma using machine learning algorithms are taken into consideration. Representative feature extraction is necessary for effective and robust diagnosis by lesion segmentation from the vicinity of usual tissue [20]. Segmentation methods can be categorized into histogram thresholding methods [21], unsupervised clustering approaches [22-26], area-based and edge-based methods [27-29], active contour methods [30], and last category is supervised segmentation methods. Support vector machine (SVM), decision tree (DT), and artificial neural network (ANN) are used for skin lesion segmentation by training the recognizers [25, 31]. Since both of these approaches rely on pixel-level features, they could not achieve satisfactory results and cannot address issues such as color imbalance, low contrast, lighting variations, hair appearance, and other artifacts in dermoscopy photos. Currently, deep learning-based approaches, in particular CNNs, have achieved significant success in segmentation and object localization issues [14-16, 32]. Extracting robust raw images and data with high-level features and hierarchical feature learning ability is the main reason for the success of CNN. There are several architectures of CNN's family for object localization and segmentation [32, 33]. Various deep CNN architectures such as U-Net[34], SegNet [33], fully convolutional neural network (FCN) [35], and DeepLab [36] have been suggested for semantic segmentation. In recent years, CNN structures advances in the semantic segmentation field have been utilized by researchers in skin lesion segmentation. For example, in 2019, Zhang et al. presented an automatic model for boundary segmentation of skin lesions by combining the features from the shallow network of FCN [37]. Ali et al. suggested skin lesions border detection using the U-Net model for the segmentation of skin lesions from dermoscopy images and edge filters [38]. Yuan et al. proposed the segmentation of skin lesion method that enhances the FCN architecture by using Jaccard distance to solve the imbalance issue between the surrounding pixels of skin lesion and validate with the ISBI 2016 and the PH2 datasets [15]. In 2020, Yanyang et al. proposed a model based on transfer learning and adversarial domain adaptation for cross-domain skin disease classification with two-phase by fine-tuning the network on two skin disease datasets [39].

III. METHODS

This study proposes a method for real-time melanoma skin lesions segmentation in real-time. One of the crucial phases is identifying the lesion location in the image, and the second phase is the segmentation of the lesion using this spatial information. The detection and segmentation processes are presented in Figure 1.

*A. F-YOLOv3 for melanoma localization*

Before starting Yolo version 3 to train, all training set data was reshaped to $512 \times 512$. During the training, the image information includes coordinates for the midpoint (x, y), height (h), and bounding box width is shown by (w) and defines the class of object to be detected. The bounding box coordinates (x2, y2) in the corner of the lower right and the corner of the upper left (x1, y1) to determine the coordinates of x and y, width (w), and height (h) shown in Figure 2.

Based on the idea of this article, there is no need to use other preprocessing phases to remove artifacts such as gel bubbles, clinical rule marks, color charts, and black frames because only the lesion area is extracted and fed into the fine-tuned segmentation network. Due to the selection of specified lesion area, many artifacts were removed in this step, and computational overhead decreased.

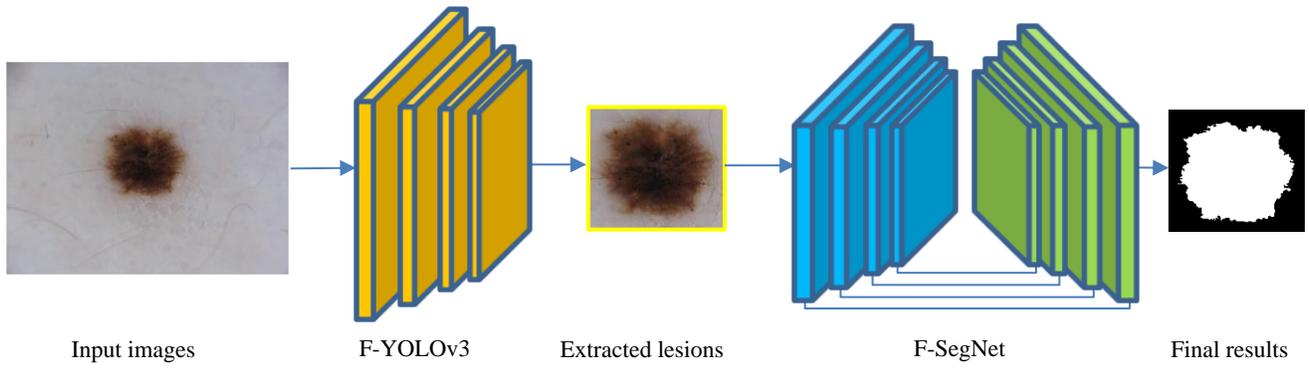

Fig 1. Melanoma lesion localization and segmentation.

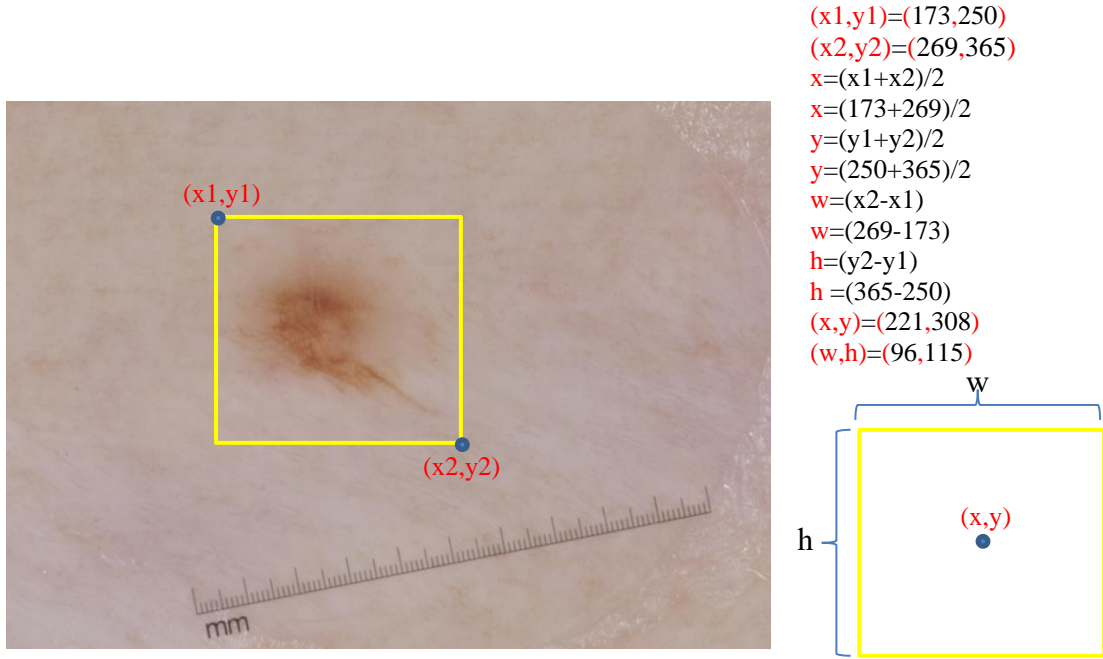

Fig 2. Skin lesion labeling for the training of F-YOLOv3

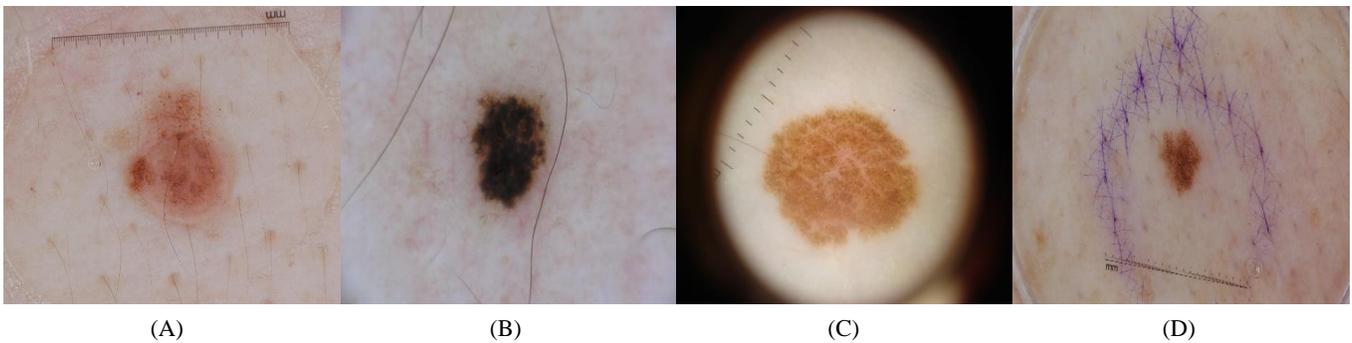

Fig 3. Skin lesion artifacts. A) clinical rule mark. B) hair. C) black frame. D) gel bubble and rule mark.

In the proposed F-YOLOv3 method, the entire image feeds into a convolutional neural network, predicting several bounding boxes with all class probabilities. The predictions are tensor with the dimension of $(S \times S) * B * (5 + C)$. The input image is split into a non-overlapped $(S \times S)$ grid cell by F-YOLOv3 responsible for their confidence scores and bounding box prediction B [32]. The confidence score is the lack or presence of an object within the bounding box.

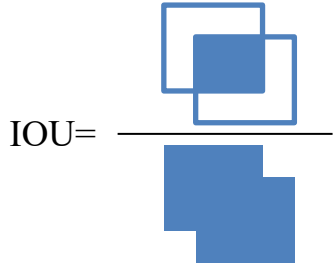

Fig 4. Representation of the IOU for measuring object detection performance [40]

The product of the intersection over union (IOU) and an object's existence probability is confidence [40].

$$confidence = \Pr(object) * IOU_{predicted}^{ground\ truth} \quad (1)$$

If the cell contains no object, the confidence score is equivalent to IOU because the confidence score is zero.

Each bounding box contains the confidence score and x, y, w, and h variables. X and y represent the bounding box center coordinates whenever h and w demonstrate the height and width values. For the class scores, there is a C variable. Each network cell makes a prediction of the probability of a conditional class $Pr(Class_i|object)$ [32]. In F-YOLOv3, the specific confidence score for each class is calculated for any box by-product of the single box confidence prediction at the test time and class probabilities as presented in Formula (1). These scores demonstrate the probability of the desired melanoma class in the box.

$$\Pr(Class_i object) * \Pr(object) * IOU_{predicted}^{ground\ truth} =$$

$$\Pr(Class_i) * IOU_{predicted}^{ground\ truth} \quad (2)$$

The GoogLeNet [41] inspired the YOLO architecture that includes the 24 convolution layers used to extract features to predict coordinates and the class probabilities output with two fully connected layers. For the last layer, Yolo utilizes a linear activation function, while YOLOv3 [32] for the other layers uses a leaky rectified linear unit (Leaky ReLU) defined as follows:

$$Leaky\ ReLU = (0 \cdot 1 * x \,.\, x) \quad (3)$$

The loss function in YOLO is the sum of the error squares that is described as:

$$loss = \lambda_{coord} \sum_{i=0}^{S^2} \sum_{j=0}^{B} 1_{ij}^{obj}[(x_i - \hat{x}_i)^2 + (y_i - \hat{y}_i)^2]$$
$$+ \lambda_{coord} \sum_{i=0}^{S^2} \sum_{j=0}^{B} 1_{ij}^{obj}\left[(\sqrt{w_i} - \sqrt{\hat{w}_i})^2\right.$$
$$\left. + (\sqrt{h_i} - \sqrt{\hat{h}_i})^2\right] + \sum_{i=0}^{S^2} \sum_{j=0}^{B} 1_{ij}^{obj}(C_i - \hat{C}_i)^2$$
$$+ \lambda_{noobj} \sum_{i=0}^{S^2} \sum_{j=0}^{B} 1_{ij}^{noobj}(C_i - \hat{C}_i)^2$$
$$+ \sum_{i=0}^{S^2} 1_{ij}^{obj} \sum_{c \in classes} (\mathcal{P}_i(c) - \hat{\mathcal{P}}_i(c))^2 \quad (4)$$

$x_i$ and $y_i$ show the bounding box center point coordinates around the lesion area, where $h_i$ and $w_i$ represent the height and width. The confidence score is the $C_i$ and $\mathcal{P}_i(c)$ which is the classification loss. The λ values are constants used to gain the loss of lesion area coordinate predictions and reduce the confidence predictions loss from bounding boxes with any objects. These parameters are set as $\lambda_{coord} = 5$ and $\lambda_{noobj} = 0.5$. Finally, $1_{ij}^{obj}$ indicates the bounding box jth predictor in cell i for its prediction is responsible and $1_i^{obj}$ represents an object in cell i. In the Yolo loss equation, the first two layers calculate the loss of localization, the confidence loss is calculated with the third layer, and the last layer measures the classification loss [32].

Skin lesion localization in three scales is presented in Figure 5. In melanoma skin lesions detection, our system trains the proposed F-YOLOv3 architecture with the ISIC 2018 dataset. Data sets were divided into 30% of the data for validation and 70% for training sets. Final detection system performance was assessed using the mentioned data set and training parameters of the proposed F-YOLOv3, as seen in the following Table 1.

Table 1. F-YOLOv3 training parameters

| Parameters | Values |
| --- | --- |
| Batch size | 32 |
| Momentum | 0.9 |
| Learning rate | 1e-3 |
| Epochs | 10000 |
| Image train size | 512×512 |
| IOU threshold value | 0.2 |
| Gradient optimizer | Adam optimizer |

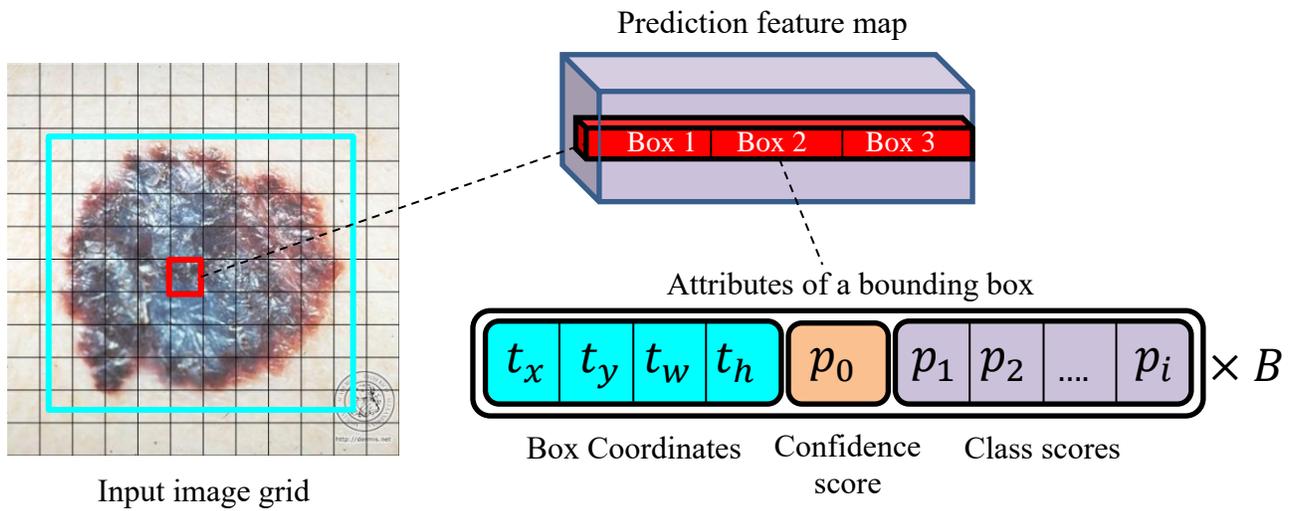

Fig 5. Melanoma with confidence predicted by F-YOLOv3. The blue frame is a bounding box with $t_x$ $t_y$ $t_w$ $t_h$ coordinates created by the red grid. $p_0$ is confidence score, and $p_i$, is the class of probability scores.

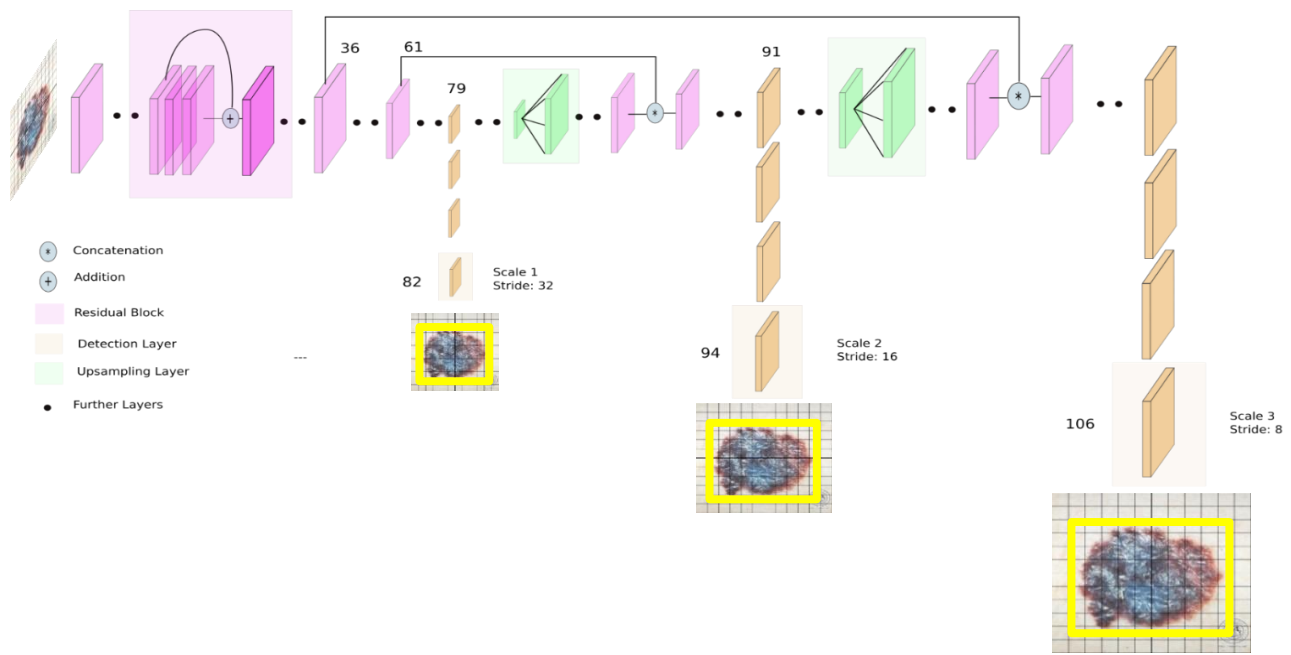

Fig 6. Melanoma skin lesion detection using the F-YOLOv3 network.

Due to the new domain of our dataset, we fine-tuned weights of YOLOv3 to our task to achieve better performance in melanoma lesion detection. In addition, we changed the last layer of YOLOv3 to detect the melanoma lesion class at the output.

As shown in Figure 7, the target domain of our proposed novel pipeline based on deep learning networks is the new domain. In the source domain, which consists of 80 classes of the COCO dataset [42], YOLOv3 was trained, and we used extracted weights for the target domain. The target domain of our proposed F-YOLOv3 is a melanoma skin lesion class that is tuned with new weights of the ISIC 2018 dataset.

Figure 8 A) shows an example of the melanoma lesion area given in the F-YOLOv3, and after calculating the bounding boxes in Figure B). Finally, the area where the lesion is entirely inside, as shown in Figure C). After this step, all extracted melanoma lesion images are prepared for feeding into F-SegNet in the next step.

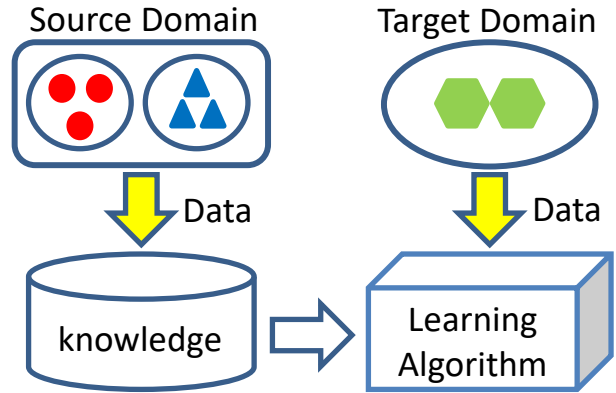

Fig 7. Illustrates the transfer learning process with the source domain from original weights and the target domain as a domain of the melanoma dataset.

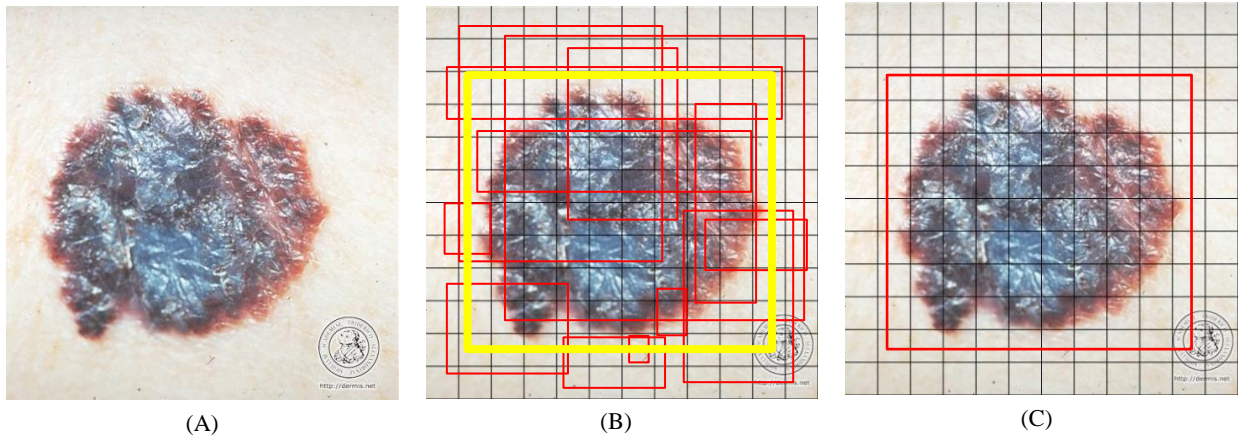

Fig 8. Progress of the proposed F-YOLOv3 to extract melanoma lesion of a clinical image. A) input images. B) bounding boxes and confidence. C) final detection.

F-YOLOv3 was trained on the ISIC 2018 dataset with the distribution specified in Table 2 with dimensions of $416 \times 416$. In training progress, we use 2594 images and corresponding ground truth.

Table 2. Original data distribution of the ISIC 2018 dataset

| Class type | Training | Validation | Test | Total |
|---|---|---|---|---|
| B | 6705 | - | - | 6705 |
| SK | 1099 | - | - | 1099 |
| BCC | 514 | - | - | 514 |
| AK | 327 | - | - | 327 |
| DF | 115 | - | - | 115 |
| VL | 142 | - | - | 142 |
| M | 1113 | - | - | 1113 |
| Total | 10,015 | 193* | 1512* | 10,015 + 1705* |

*Not available

The results of the F-YOLOv3 architecture evaluation on the ISIC 2018 dataset are listed in Table 3. 53 convolutional layers, 246 million parameters, which is the result of multiplying the length and width of the filter by the number of filters in the previous layer plus one and finally multiplying by the number of filters, detection speed of 35 frames per second and average accuracy of 96% in Table 3 is given.

Table 3. performance of F-YOLOv3

| | Number of Convolutional Layers | Parameter of Weight(M) | FPS | mAP |
|---|---|---|---|---|
| F-YOLOv3 | 53 | 246 | 35 | 0.96 |

The automatic pipeline for melanoma skin lesions segmentation is based on the proposed F-YOLOv3 and F-SegNet methods presented in this study. In the proposed pipeline, the lesion location is identified in the input images. The F-YOLOv3 is tuned for a melanoma class that will specify the location of melanoma skin lesions at the output. The detected kernel size is applied to the feature map as 1×1. Finally, B specifies the number of bounding boxes per cell, where the detection kernels shape will be $(5 + C) * B * (1 \times 1)$. C is the class size, and constant 5 indicates the number of bounding box parameters. According to the formula, we have a class, so the filters will be $=3\times (5 + classes)$. After F-YOLOv3 training, a frame is automatically formed on the melanoma skin lesion, and it is the starting point for segmentation. The specified lesion location is then given to the segmentation network. This approach avoids wasting time searching for the location of the lesion and the high computational load and allows only the identified location to be processed.

*B. F-SegNet for melanoma segmentation*

The ISIC 2018 data set, after passing in F-YOLOv3, prepared images are entered into F-SegNet for training with $360 \times 480$ dimensions and according to the training parameters in Table 4. Finally, we will have the segmentation image at the network output. Fine-tuned SegNet [33] architecture for melanoma class has been presented in Figure 9.

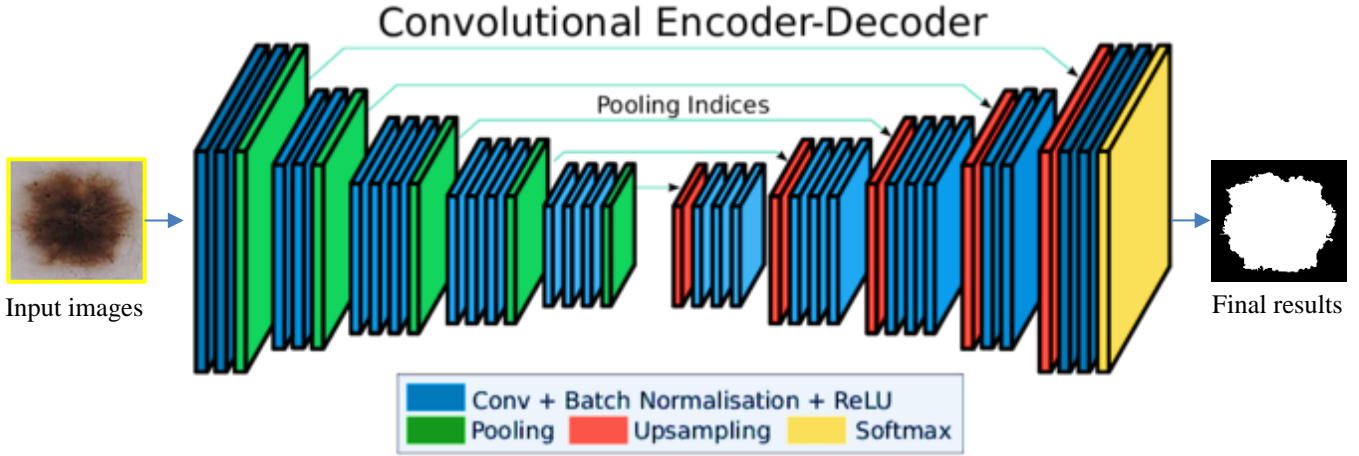

Fig 9. F-SegNet layers with the input image and segmented output melanoma class.

The final step includes training the F-SegNet method with the images from the F-YOLOv3 output. In particular, we fine-tuned the SegNet [33] model with the images of localized melanoma lesions. SegNet has a symmetrical encoder-decoder structure which allows learning significantly well and training on urban scenes. The decoder network uses upsampling layers that map the low-resolution representation to pixel-wise predictions of the desired class, and the encoder network provides a low-resolution representation.

Table 4: F-SegNet learning parameters

| Parameters | Values |
|---|---|
| Batch size | 32 |
| Optimizer | Adam |
| Learning rate | 0.001 |
| Epochs | 200 |

In the second step of our learning-based pipeline in this paper, for the segmentation section, we performed some modifications on SegNet for the experiments and the loss function using the median frequency balancing (MFB) shown in equation (5). Using the cross-entropy loss, we train the F-SegNet. By adding the MFB (ϕ) into this function as follows:

$$J(\theta) = -\frac{1}{m}\left[\sum_{i=1}^{m}\phi_y(i)[y^{(i)}\log\hat{y}^{(i)} + (1-y^{(i)})\log(1-\hat{y}^{(i)})]\right] \quad (5)$$

Where the label is $y^{(i)}$, $m$ is the labeled pixels, and the prediction of output is $\hat{y}^{(i)}$. On our data, this function improves output with the class frequency average ratio, weighted melanoma class in the loss function, and calculated on the whole training collection separated by the class frequency.

Figure 10 is the original image of the dataset, Figure 11 is the result before enhancing and tuning the segmentation model, and Figure 12 shows the output image of the proposed F-SegNet after using the proposed F-YOLOv3. So we can observe that our results are outstanding on the unseen test set, which suggests that the model is robust and well-performing too. If we calculate the difference between predicted output before and after fine-tuning, the mean difference is 8.23% in accuracy, which means that fine-tuning is a crucial step for similar tasks.

In addition, because of the relation between TP, TN, FP, and FN, other parameters presented in table 6 will be improved by fine-tuning.

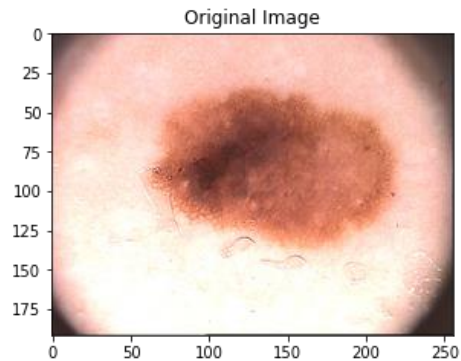
Fig 10. Original image

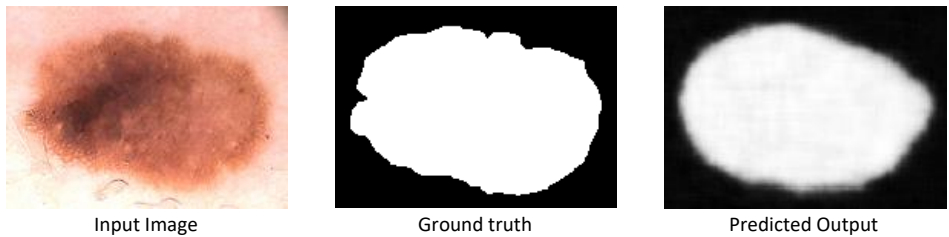
Fig 11. Before fine-tuning the segmentation model

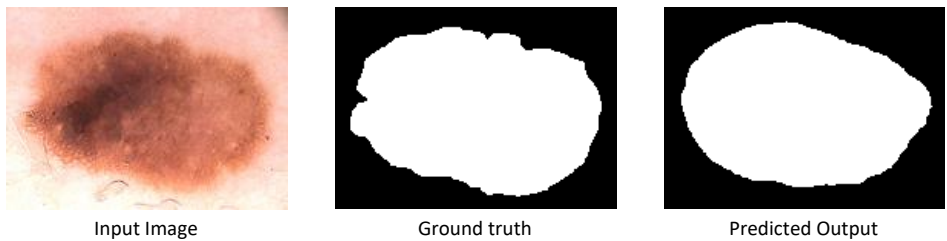
Fig 12. After fine-tuning the segmentation model

## IV. RESULTS

According to this article, the melanoma dataset evaluates in a pipeline with two-step. The first step investigated the detection function of melanoma skin lesion's location from F-YOLOv3 retraining. The lesion location in the image is critical for obtaining more precise segmentation results. Using the IOU criterion, the model localization efficiency was validated.

If the IOU of the lesions is greater than 80%, the location of the identified lesion is considered correct. The proposed approach is further evaluated in the second stage using the following performance criteria:
- Dice similarity coefficient
- Sensitivity
- Jaccard Index
- Accuracy

Sensitivity (SE) indicates the number of correctly segmented lesion pixels. Dice similarity coefficient (D) is an evaluation criterion used to evaluate segmented lesions and the likeness of a correctly interpreted basis. An evaluation criterion for the intersection ratio between the correct basis masks and the obtained segmentation outcomes is the Jaccard index (J). The primary distinction between IOU and J is that IOU is used for localization, and its borders are rectangular. Still, J is used for segmentation, and the segmentation boundaries may be ruleless. Mean average precision (mAP) is used to evaluate the object localization model. The mAP computes a score by comparing the ground-truth bounding box to the observed box. The model's detections are reliable when the score is higher. Finally, accuracy indicates the overall pixel-based segmentation metric. The following equations are used to compute all of the above evaluation criteria:

$$IOU = \frac{area\ of\ Overlap}{area\ of\ Union} \quad (6)$$

$$Specificity = \frac{TN}{TN+FP} \quad (7)$$

$$Sensitivity = \frac{TP}{TP+FN} \quad (8)$$

$$Dice\ coefficient = \frac{2*TP}{(2*TP)+FP+FN} \quad (9)$$

$$Jaccard = \frac{TP}{TP+FN+FP} \quad (10)$$

$$Accuracy = \frac{TP+TN}{TP+TN+FP+FN} \quad (11)$$

The letters FP, TP, TN, and FN represent false-positive, true positive, true negative, and false negative. The image's lesion pixels are regarded as TP if correctly segmented, otherwise known as FN. In contrast, if the prediction is a non-lesion pixel, the pixels with no TN image lesion are considered. Otherwise, they have been known as FP.

The F-YOLOv3 is tuned to make the melanoma lesion prediction, and the neck section incorporates Feature Pyramid Network (FPN) to collect the DarkNet-53 backbone network's deep feature maps. Because of three different pyramid levels, this collection of feature maps detects smaller melanoma skin lesions. A comparison of our F-YOLOv3 with three families of YOLO evaluated by the ISIC 2018 melanoma dataset can be seen in Table 5. Results show that F-YOLOv3 achieved higher performance with 0.96 in the mAP metric.

As shown in Table 5, the YOLO family evolution varies in technical considerations of the detector's backbone structure, neck, and head, making the newer version more efficient than the prior. The primary variant of the YOLOv1 [43] architecture supports end-to-end training and uses an end-to-end differentiable backbone 24-layered network. The backbone model of YOLOv2 [44] is DarkNet-19, with five max-pooling layers and 19 convolution layers. The head and neck sections in YOLOv2 and YOLOv1 relied on non-max suppression and Softmax for localization and prediction of melanoma skin lesions.

The result of melanoma lesion localization by F-YOLOv3 from test samples of the ISIC 2018 dataset is shown in Figure 13 with different artifacts. In addition, before tuning the model, the region of melanoma will localize but have lower accuracy and more area to detect, which will increase the computational time. Furthermore, it will prevent us from using and implementing the model on the less expensive hardware. To have a suitable size for the input of the segmentation phase, we added zero padding to each extracted image.

Table 5. comparison of YOLO family for melanoma localization

| YOLO | Backbone | Neck | Head | mAP |
|---|---|---|---|---|
| YOLOv1 | 24-layered CNN | Non-max suppression | Softmax | 0.37 |
| YOLOv2 | DarkNet-19 | Non-max suppression | Softmax | 0.83 |
| YOLOv3 | DarkNet-53 | Feature pyramid network | Logistic regression | 0.93 |
| **Our F-YOLOv3** | DarkNet-53 | Feature pyramid network | Logistic regression | **0.96** |

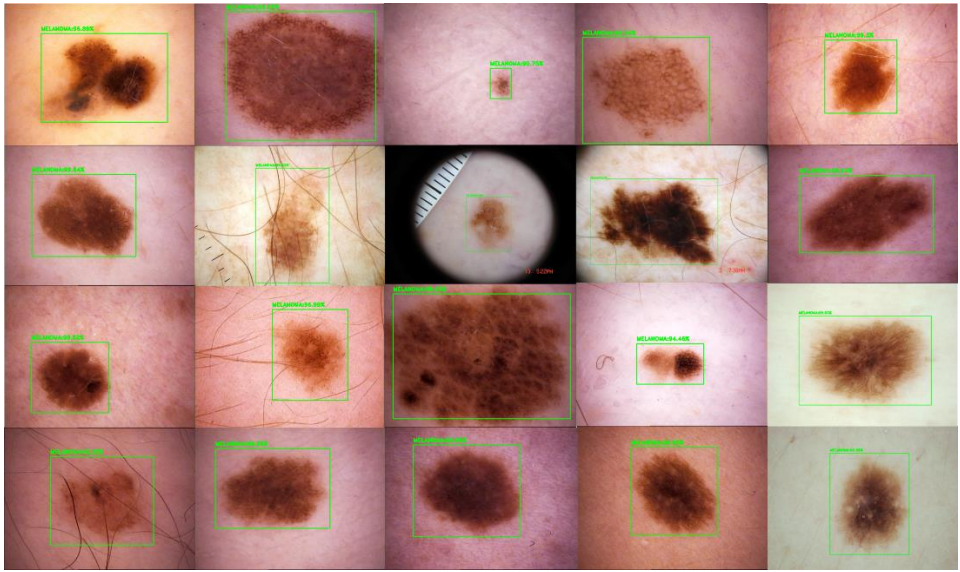

Fig 13. Test samples of the ISIC 2018 dataset localization step after tuning the F-YOLOv3.

Table 6. F-SegNet performance comparison with similar melanoma segmentation task on ISIC 2018. All values are in percentage.

| Method | A | D | J | SE |
|---|---|---|---|---|
| U-Net [34] | 91.87 | 81.67 | 72.68 | 90.43 |
| U-Net++ [45] | 92.49 | 84.37 | 74.35 | 88.89 |
| FCN [46] | 90.11 | 78.61 | 70.13 | 89.66 |
| R2Net [47] | 91.72 | 82.71 | 75.11 | 91.22 |
| BCU-Net [48] | 93.37 | 86.37 | 76.65 | 92.72 |
| FocusNetAlpha | 94.47 | 90.14 | 82.71 | 94.71 |
| Rehman et al. [49] | 94.2 | 93.1 | 91.87 | 93.2 |
| Zhang et al. [50] | - | 83.8 | 74.4 | - |
| **Our F-SegNet** | **95.16** | 92.81 | 86.2 | 94.62 |

From the results reported in Table 6, we compare our model with similar segmentation methods on the ISIC 2018 dataset. Also, the specificity in our work is 93.2%. We observe a considerable increase in accuracy, dice similarity coefficient, and Jaccard index in our F-SegNet compared to cutting-edge approaches.

## V. DISCUSSION

The F-YOLOv3 and the F-SegNet for melanoma lesions detection and segmentation have been presented. We compared our F-YOLOv3 method with the well-known detector model family, as represented in Table 5, with the membership of YOLOv1 [43], YOLOV2 [44], and YOLOv3 [32]. After fine-tuning some blocks in the YOLOv3 structure on ISIC 2018 dataset, we achieved better performance in the mAP reported in Table 5. Since the trained F-YOLOv3 method only detects desirable areas for segmentation with high mAP localized melanoma lesions.

To further evaluate the performance of our the F-SegNet with cutting-edge segmentation methods, U-Net [34], U-Net++[45], FCN [46], R2Net [47], BCU-Net [48], FoucusNetAlpha [51] produce a lower value of accuracy and dice coefficient than our method. Architecture types and parameter values in these networks are different. At the same time, our proposed F-SegNet approach segments melanoma lesions based on pixel-wise segmentation.

The reason behind good performance in the F-YOLOv3 is predicting melanoma lesion area and being capable of defeating all the various artifacts. In addition to being used to diagnose melanoma, F-YOLOv3 can also be used for medical images of non-melanoma skin lesions with natural artifacts. Challenging of accurate melanoma localization because of the visual resemblance to ordinary melanoma lesion non-uniformity, and the border is irregular. The wide range in color and melanoma lesion texture makes it difficult for the dermatologist to correct melanoma prediction. Deep learning-based tools like F-YOLOv3 and F-SegNet enable the CAD solutions to precisely melanoma localization and segmentation, thus improving patients' life expectancy. We can improve the models to better results and work on fine-tuning the specific layers to achieve better performances.

## VI. Conclusion

This article presented a novel scheme for detecting and segmentation melanoma skin lesions with F-YOLOv3 and F-SegNet models. Against the prior deep model-based segmentation approaches, our approach consists of two-phase: skin lesion localization and segmentation of localized lesions. In addition, we have more computational steps without fine-tuning because unnecessary regions will feed into the model. Our techniques precisely and efficiently segment and detect in comparison to cutting-edge models. We evaluated the proposed methods using ISIC 2018 well-known dataset. Also, the suggested ways can be utilized in other medical image detection and segmentation issues, such as MR images. To improve this study, we can apply more fine-tuning processes in both phases to overcome recent manuscripts.